\documentclass[12pt]{article} 

\usepackage{hyperref}
\usepackage{url}
\usepackage{setspace}
\usepackage{amsmath,amssymb,amscd,braket}
\usepackage{amsfonts}
\usepackage{color}
\usepackage{graphicx}
\usepackage{cite}
\usepackage{amsthm}

\numberwithin{equation}{section}

 \newcommand{\bea}{\begin{eqnarray}}
\newcommand{\eea}{\end{eqnarray}}
\newcommand{\be}{\begin{equation}}
\newcommand{\ee}{\end{equation}}
\newcommand{\ba}{\begin{align}}
\newcommand{\ea}{\end{align}}
\newtheorem{result}{Landscape Result}

\newcommand\rref[1]{(\ref{#1})}

\newlength{\slength}
\renewcommand{\title}[1]{\vbox{\center\LARGE{#1}}\vspace{5mm}}
\renewcommand{\author}[1]{\vbox{\center#1}\vspace{5mm}}
\newcommand{\address}[1]{\vbox{\center\footnotesize\em#1}}
\newcommand{\email}[1]{\vbox{\center\footnotesize\tt#1}\vspace{5mm}}

\begin{document}

\begin{titlepage}

\begin{center}

\hfill \\
\hfill \\
\vskip 1cm

\title{A measure on the space of CFTs and pure 3D gravity}

\author{Alexandre Belin$^{a,b,c}$, Alexander Maloney$^{d,e,f}$, Florian Seefeld$^{c,g,h}$
}

\address{
${}^a$Dipartimento di Fisica, Universit\`a di Milano - Bicocca \\
I-20126 Milano, Italy

\vspace{1em}
${}^b$INFN, sezione di Milano-Bicocca, I-20126 Milano, Italy

\vspace{1em}
${}^c$ Institute of Physics, Ecole Polytechnique F\'ed\'erale de Lausanne, \\ CH-1015 Lausanne, Switzerland

\vspace{1em}
${}^d${Department of Physics, Syracuse University, Syracuse, NY}

\vspace{1em}
${}^e${Institute for Quantum and Information Sciences, Syracuse University, Syracuse, NY}

\vspace{1em}
${}^f${Department of Physics, McGill University, Montreal, Canada}

\vspace{1em}
${}^g$ Centre for Quantum Mathematics, University of Southern Denmark, \\ 5230 Odense M,
Denmark

\vspace{1em}
${}^h$ Department of Mathematics and Computer Science, University of Southern Denmark, \\ 5230 Odense M,
Denmark}

\email{alexandre.belin@unimib.it, alex.maloney@mcgill.ca, fseefeld@imada.sdu.dk}

\end{center}

\abstract{We define a normalizable measure on subspaces of two-dimensional conformal field theories, which we interpret as a maximum ignorance ensemble.  These subspaces are specified by a range of allowed central charges and a minimum gap on the scaling dimension of operators. We test whether pure quantum gravity in AdS$_3$ is dual to the average over this ensemble. We find a negative answer, which implies that CFTs with a  primary gap of order the central charge are highly atypical in our ensemble. We provide evidence that more generally, holographic CFTs are atypical in the space of all CFTs by finding similar results for permutation orbifolds: subgroups of $S_N$ with a good large $N$ limit are very sparse in the space of all subgroups. Along the way, we derive several new results on the space of CFTs. Notably we derive an upper bound on the spacing in central charge between CFTs, which is doubly exponentially small in the large central charge limit.

}

\vfill

\end{titlepage}

\eject

\tableofcontents

\section{Introduction}

The study of low-dimensional models of gravity has shed light on some of the most mysterious aspects involving quantum gravity, most notably the black hole information paradox \cite{Penington:2019npb,Almheiri:2019psf,Penington:2019kki,Almheiri:2019qdq}. Most of the recent progress has followed from an in-depth study of Jackiw-Teitelboim (JT) gravity in two-dimensional Anti-de Sitter (AdS) space. JT gravity is a simple toy-model for quantum gravity, which has no local propagating degrees of freedom. Nevertheless, the gravitational theory remains non-trivial due to boundary excitations which give rise to the Schwarzian theory and to the inclusion of non-trivial topologies contributing to the gravitational path integral. Remarkably, the gravitational path integral can be explicitly computed and it agrees with a particular matrix model in a double-scaled limit \cite{Saad:2019lba}. 

This outcome should be contrasted with the standard incarnations of the AdS/CFT correspondence, most notably that of $\mathcal{N}=4$ SYM and type IIB string theory on $AdS_5\times S^5$. The foundation of the correspondence relies on the equality of partition functions between the bulk and boundary theories
\be
Z_{\text{AdS}}=Z_{\text{CFT}} \,.
\ee
For JT gravity, the prescription is modified and reads
\be \label{averageintro}
Z_{\text{AdS}}= \int dH P(H) \ Z_{\text{CFT}(H)} \,.
\ee
The gravitational path integral is no longer dual to the partition function of a quantum mechanical system with definite Hamiltonian, but rather to the ensemble average over some probability distribution $P(H)$ of quantum mechanical Hamiltonians $H$. This implies that the boundary dual of pure quantum gravity in AdS$_2$ is not quantum mechanics, which requires a fixed Hamiltonian.

The appearance of an ensemble average shed lights on the contribution of Euclidean wormholes in the gravitational path integral. As first observed in \cite{Maldacena:2004rf}, the contribution of Euclidean wormholes in the gravitational path integral spoils the factorization of CFT partition functions on disconnected Euclidean boundaries. If we consider boundary conditions for the gravitational path integral given by the union of two components $\mathcal{B}_1 \cup\mathcal{B}_2 $, we find
\be
Z_{\text{AdS}}(\mathcal{B}_1 \cup \mathcal{B}_2 ) \neq Z_{\text{AdS}}(\mathcal{B}_1)  Z_{\text{AdS}}( \mathcal{B}_2 ) \,.
\ee
This factorization puzzle is a feature rather than a bug, if the boundary theory is indeed an ensemble-average. This follows from the fact that the integral over Hamiltonians correlates the two disconnected boundaries
\be
Z_{\text{AdS}}(\mathcal{B}_1 \cup \mathcal{B}_2 ) = \int dH P(H) \ Z_{\text{CFT}(H)}(\mathcal{B}_1)Z_{\text{CFT}(H)}(\mathcal{B}_2) \,.
\ee

But the factorization puzzle must also be addressed in higher dimensional AdS/CFT, as Euclidean wormholes exists in arbitrary dimensions \cite{Maldacena:2004rf,Marolf:2021kjc,Maloney:2025tnn}. There however, it is far from obvious what type of ensemble one should be considering, or if there is any ensemble averaging whatsoever in the first place. After all, one can study $\mathcal{N}=4$ SYM at fixed $N$ and $g_{YM}$ where explicitly, no averaging is involved. Understanding the implications learned from JT gravity to higher dimensions remains an open problem. The goal of this paper is to make progress on this front.

In terms of higher dimensional holography, the natural first step is three-dimensional gravity. 3D quantum gravity in AdS is the perfect "halfway-point``between JT gravity and a top-down theory like $\mathcal{N}=4$ SYM. A theory of pure 3D quantum gravity shares many similarities with JT gravity: perturbative UV-completeness, no local propagating degrees of freedom yet an interesting spectrum of boundary excitations (Virasoro generators replace the Schwarzian mode of JT gravity), and a special role played by constant negative curvature metrics.\footnote{There is an important distinction between JT gravity and 3D gravity: in JT gravity, the dilaton enforces the fact that the full off-shell path integral localizes on constant negative curvature surfaces. In 3D gravity, constant negative curvature metrics are the solutions to the equations of motion while the full path integral would receive contributions from off-shell geometries.} 

However, 3D gravity drastically differs from JT gravity at the level of its dynamical constraints. Indeed, the boundary dual of quantum gravity in AdS$_3$ should be a two-dimensional CFT (or perhaps an ensemble thereof), but CFTs are highly constrained by modular invariance and crossing symmetry. This is very different from quantum mechanical Hamiltonians, where any hermitian matrix is as good as any other at the level of defining a consistent theory. We can thus define any probability measure over hermitian matrices and this defines a probability measure over consistent theories. One of the major difficulties in trying to build ensemble averages in higher dimensions is dealing with the constraints that CFTs should satisfy: CFTs should be crossing-symmetric and modular invariant. Averaging over the spectrum of the dilation operator without extreme care would badly violate these constraints and lead to undesirable pathologies.

There is accumulating evidence that the dual of pure three-dimensional gravity in AdS$_3$ is dual to some type of ensemble average. Maloney, Witten and Keller \cite{Maloney:2007ud,Keller:2014xba} computed the thermal partition function of 3D gravity by summing over all smooth solutions to the equations of motion. Extracting the density of states from this calculation yielded two undesirable properties. First, the density of states was a continuous function of energy. Second, the density of states was negative in some regions of conformal weights, see also \cite{Benjamin:2019stq}. Both properties were seen as problematic for a direct interpretation in terms of a unitary compact conformal field theory dual to 3D gravity. The negativity of the density of states is still unresolved, but there are several different ways to cure the negativity: one can either add off-shell contributions to the path integral \cite{Benjamin:2020mfz,Maxfield:2020ale}, or add new non-black hole states \cite{DiUbaldo:2023hkc}. The continuum of states, on the other hand, has a natural interpretation in hindsight: 3D gravity is dual to an ensemble average over quantum systems. For smooth probability distributions, performing an ensemble of the type \eqref{averageintro} will always yield a continuous density of states. Such an average also predicts the breakdown of factorization, and the appearance of Euclidean wormholes. And indeed, the ensemble interpretation is reinforced by the matching of many on-shell wormholes in 3D gravity with an ensemble of 2d CFTs \cite{Belin:2020hea,Chandra:2022bqq,Collier:2024mgv,deBoer:2024mqg,Chandra:2025fef}.

The construction of the ensemble can be approached from multiple view-points. One can either define a set of moments for the CFT data (scaling dimensions, spin, and OPE coefficients of the CFT operators), and work in a moment expansion. This was the approach taken in \cite{Chandra:2022bqq}, and the leading order statistics for the OPE coefficients is Gaussian. This reproduces many on-shell Euclidean wormholes of the bulk theory. The issue is that the moments of OPE coefficients are highly non-Gaussian \cite{Belin:2021ryy,Anous:2021caj}, and it quickly becomes hard to track the moments of the CFT data. Alternatively, one can follow the approach of \cite{Belin:2023efa} and define directly a matrix-tensor model whose potential is obtained from the crossing equation, which guarantees that all the moments are correctly encoded. The tensor model then generates, through its Feynman diagrams, the sum over bulk topologies \cite{Jafferis:2024jkb}, with a contribution for each topology that agrees with the bulk partition function on a given topology (even at finite central charge), formulated most elegantly as a Virasoro TQFT \cite{Collier:2023fwi,Collier:2024mgv} or conformal Turaev-Viro theory \cite{Hartman:2025cyj,Hartman:2025ula}. The tensor model also encodes non-trivial statistics for the density of states, which in the bulk come from an off-shell wormhole with torus boundaries \cite{Cotler:2020ugk} (see also \cite{Boruch:2025ilr}).

These developments hint that a precise formulation of the CFT dual to pure quantum gravity in AdS$_3$ is close. However, there are important subtleties in either formulation of the ensemble which are not currently understood. In the moment version of the ensemble, the issue lies in the fact that it is unclear whether the set of moments form a consistent ensemble: moments of a probability distribution need to satisfy an infinite number of consistency conditions (the positive semi-definiteness of the Hankel matrix in the Hamburger moment problem), and it is unclear whether this is indeed the case. This is connected to tails in the probability distribution of the CFT data, which one is trying to squeeze to satisfy the crossing equations exactly. In the tensor model, the subtleties arise due a certain number of scaling limits that need to be taken to reproduce the gravity limit. This is most likely a triple scaling limit, but the precise nature of this limit is not currently understood. In either case, the subtlety comes from trying to define a probability distribution that in the right limit, imposes crossing exactly, since any bulk computation is, almost by definition, exactly crossing invariant. 

The goal of this paper is to bypass these problems, and to define an average directly over the space of CFTs. We will thus consider the following ensemble average
\be
\overline{ \cdot}\equiv \int_{C(c_0,\epsilon_c)} d\mu  \cdot \,, \label{ensembleintro}
\ee
where $C (c_0,\epsilon_c)$ represents the space of CFTs in some window of size $\epsilon_c$ around a mean central charge $c_0$ and $\cdot$ is any observable that can be defined over the ensemble (for example a torus partition function). This requires defining a normalizable measure on the space of CFTs, which will be one of our results. We will then investigate the extent to which such an ensemble can be dual to pure gravity in AdS$_3$.\footnote{Note that for rational CFTs, a sum over topologies in the dual TQFT can produce a measure over the space of CFTs with that central charge \cite{Dymarsky:2024frx}. It would be interesting to understand the connection with this work.}

It is important to note that we have traded the issue of the precise definition of an ensemble of CFT data with another (potentially harder) problem: that of understanding the space of CFTs at a given central charge, an unsolved problem even in $d=2$. Our understanding of this space is limited by the set of CFTs we know how to construct, but it is worth noting that these may be highly atypical in the space of CFTs. The best example of this is the Narain CFTs (free bosons on a lattice). Even with a lot of determination, it was impossible to construct families of lattices of increasing dimension such that the scaling dimension of the lightest primary scaled linearly with $c$. This is the equivalent to finding a 2d CFT with only Virasoro symmetry and a large gap of order $c$ to the first primary operator, but with Virasoro replaced by the chiral algebra $U(1)^c$. Averaging over the Narain ensemble produced an average gap of order $c$ \cite{Maloney:2020nni,Afkhami-Jeddi:2020ezh}, with a variance exponentially small in $c$ \cite{Collier:2021rsn}.\footnote{It is worth mentioning that both the mean and variance results are obtained by computing the partition function directly, and then manipulating it to extract properties about the gap. This is an annealed free energy type of computation. A direct extraction of the typical gap would require a log and would be similar to a quenched-free energy type of computation. It would be interesting to see if the two results agree in scaling.} This means that the typical lattice is actually of order $c$, even if we cannot construct a single theory with such a gap. This shows that the space of what we currently know how to construct can be very different from a typical theory and that sometimes, it is simpler to describe the average over some ensemble, rather than trying to find/construct typical members of it.

Extrapolating this idea to the space of all CFTs, it is conceivable that a typical CFT at large central charge has a gap of order the central charge, even if we cannot currently construct a single one. Similar statements apply to the chiral algebra, it also appears conceivable that a typical CFT at central charge $c$ has only Virasoro symmetry as its chiral algebra. This would resonate well with the situation for Random Matrix Theory in quantum mechanics, where Hamiltonians with extra symmetry certainly exist, but they are highly atypical within the space of all Hamiltonians. If these statements were true, then the ensemble \eqref{ensembleintro} we will construct in this paper would produce 3D gravity as its dual bulk description.

\subsection{Summary of results}

We define a normalizable probability measure over the space of CFTs. The measure is defined as follows

\be
\int_{C(c_0,\epsilon_c,\Delta_{\min})} d\mu \equiv \frac{1}{\mathcal{N}} \left( \sum_{\mathcal{C}_i}+\sum_{\mathcal{M}_j}  \int_{\mathcal{M}_j} d\mu_j^Z \right) \,.
\ee
The details of all the various parameters that enter in this definition are given below \eqref{measuresec2}, but the main ingredients are the following. The measure is specified by three parameters: the first two parameters define a window of central charges centered at $c_0$ and of width $2\epsilon_c$. The third parameter defines a minimal gap of scaling dimensions, such that the average is taken only over CFTs where all operators have scaling dimension $\Delta> \Delta_{\min}$. This third parameter should be viewed as a regulator, since the measure we define would not be normalizable without it. The probability measure is then given by summing over all isolated CFTs with weight one and summing over all disconnected conformal manifolds with weight one over the Zamolodchikov volume, while integrating each conformal manifold with the Zamolodchikov metric. This measure treats all isolated CFTs democratically, and weighs conformal manifolds with data only from the Zamolodchikov metric, the most natural measure on the space of continuously connected CFTs. As such, it can be viewed as the maximum ignorance ensemble on the space of CFTs.

The fact that this measure is normalizable relies on a conjecture due to Kontsevich and Soibelman \cite{Kontsevich:2000yf} (discussed also by Douglas and Acharya \cite{Acharya:2006zw}), which draws inspiration from the study of Calabi-Yau sigma models. The conjecture states precisely that the space of CFTs below a given central charge is finite, and the only sources of divergence that could occur come from singular points of the conformal manifold which cause a blowup of the Zamolodchikov metric, or accumulation points of isolated CFTs. Both are expected to be correlated with towers of scaling dimensions becoming light. Here, finite means finite number of disconnected components and each component which has a conformal manifold has finite Zamolodchikov volume. We will illustrate these divergences with examples both for isolated CFTs and for conformal manifolds. This conjecture follows similar ideas to those of the swampland program and the distance conjecture, in particular its CFT incarnation \cite{Perlmutter:2020buo,Baume:2023msm,Ooguri:2024ofs}. 

Given this measure on the space of CFTs, we test whether pure quantum gravity could result from the average over all CFTs using this measure, in the limit of large central charge ($c_0\to\infty$). Unfortunately, we find that the answer is negative, at least in the limit where the regulator $\Delta_{\min}$ is removed. The problem comes from the pollution of the space of CFTs by taking tensor products of low-$c$ CFTs, like the minimal models. These products give infinite number of CFTs (that we will call accumulation points) on an extremely dense grid of central charge values. This divergence is cut-off by the regulator, but given that we assume the number of CFTs to be finite at any given cutoff, this implies that the number of pure gravity like CFTs (i.e. gap of order $c$ in primary dimensions) must also be finite. As such, at any fixed $c$, there always exists a finite $\Delta_{\min}$ at which the products of minimal models dominate over pure gravity like theories.

While this is disappointing for the purposes of 3D gravity,\footnote{We comment on possible work-arounds in the discussion section.} our venture through the space of CFTs yields interesting results on the nature of this space, as we now summarize.

\subsubsection*{Results on the space of CFTs}

We present several new results on the space of CFTs:
\begin{enumerate}
\item The space of CFTs is extremely dense in central charge. At large central charge, the difference between two values of central charges where CFTs live is\footnote{This is in fact the difference in central charge between two values at which live an infinite number of CFTs. If we want to keep $\Delta_{\min}$ finite, this result is valid provided we scale $\Delta_{\min} \sim e^{-e^{c}}$. At fixed $\Delta_{\min}$ and large $c$, the spacing would eventually becomes constant. }
\be
\delta c \lesssim e^{-e^{c}} \,.
\ee
To the best of our knowledge, this is the first place where a doubly exponential behaviour in central charge appears. Doubly exponential effects are the hallmark of the late times physics of black holes. It would be interesting to understand the significance of this result for quantum gravity in AdS$_3$. Doubly non-perturbative results in $c$ are also expected in the tensor model \cite{Belin:2023efa}, where they would encode information about the space of actual CFTs.

\item We present non-trivial accumulation points of irrational CFTs. These are accumulation points of CFTs that have conformal manifolds on which there are irrational points. By non-trivial, we mean that they are not obtained by a tensor product of one isolated irrational theory, tensored with an accumulation point of rational ones. The first example appears at central charge $c=9$, and comes from taking the symmetric orbifold of three $\mathcal{N}=(2,2)$ minimal models. Such theories accumulate in the large level limit to $c=9$, and each of these theories posses marginal operators that deform away from the orbifold point and make the CFT irrational.
\item We construct conformal manifolds of bounded central charge but of unbounded dimensionality. The simplest example appears below $c=6$, and comes from taking the product of two $\mathcal{N}=(2,2)$ minimal models. In the large level limit, the central charge remains bounded (and goes to six), while the dimensionality of the conformal manifold scales linearly with the level. Similar results with higher polynomial growth exist at any $c=3l$ with $l>2$.
\item Permutation orbifolds by a subgroup of $S_N$ with holographic properties are highly atypical in the space of all possible subgroups of $S_N$. In particular, the probability that a subgroup leads to a convergent spectrum in the large $N$ limit with a single stress tensor is more than exponentially suppressed and scales as\footnote{Similar results showing that Narain CFTs with holographic properties are sparse in the space of lattices will be presented in \cite{maloneytoappear}.}
\be
p(N) \sim e^{-N^2}
\ee

\end{enumerate}

The rest of the paper is organized as follows. In section \ref{sec:measure}, we introduce a normalizable measure on the space of CFTs and discuss various properties on the space of two-dimensional CFTs. We discuss the implications for pure 3D gravity in section \ref{sec:gravinter}. In section \ref{sec:permutation}, we discuss permutation orbifolds and a measure on the space of subgroups of $S_N$, and discuss the probability that a subgroup has a nice large $N$ limit. Finally, we conclude with a discussion in section \ref{sec:discussion}, in particular on the fate of 3D quantum gravity.

\section{A measure on the space of CFTs \label{sec:measure} }

In this section, we study the space of all CFTs with central charge $c$ and define a probability measure on that space. Our goal is to define a maximum ignorance ensemble of CFTs, and understand what the properties of a typical CFT drawn from that probability measure are.

We first start by noting that the space of CFTs is most likely a discrete space in terms of the central charge. In AdS, this means that the cosmological constant in AdS$_3$ is quantized. While there is no complete proof of this statement, it follows from the $c$-theorem that one cannot continuously change the value of the central charge by deforming a CFT with a relevant operator. The only possibility is that there are continuous families of CFTs labeled by parameters that do not correspond to deformations of the action by a local operator. There are examples of these: generalized free fields (GFF), Liouville CFT, etc. But they all have some sort of pathology (no stress-tensor and no modular invariance for the GFF, no normalizable vacuum for Liouville). To be on the safe side of this issue, we will consider the space of CFTs in a small window of central charges. We define a mean value of the central charge $c_0$, and consider the central charge window $c\in \left[c_{0}-\epsilon_c,c_{0}+\epsilon_c\right]$. We shall call this space $C(c_{0},\epsilon_c)$. The consideration of this window will guarantee that we always have CFTs in the space under consideration. We will soon see that the window size $\epsilon_c$ can be taken to be extremely small while still containing many CFTs.

We now want to define a probability measure on this region of CFT space, and understand what a typical draw looks like given this measure. From the get go, it is important to distinguish two classes of CFTs: isolated CFTs with no exactly marginal operators, and continuous families of CFTs living on conformal manifolds.\footnote{If there were continuous families of CFTs that are not related by local operators, and also do not suffer from pathologies, they would need to be taken into account here.} 

Within a given family of CFTs living on the same conformal manifold, there is a natural measure induced by the Zamolodchikov metric. However, there is no standard measure on the space of isolated CFTs, or no standard measure to compare isolated CFTs to families living on conformal manifolds.\footnote{See \cite{Douglas:2010ic,Benjamin:2023chz} for a discussion of this question along with various propositions for measures on the space of all (even disconnected) CFTs.} Since we are looking for a maximum ignorance ensemble, the most natural measure on the space of CFTs is to weigh evenly all isolated CFTs\footnote{Another possibility would be to consider as measure for individual CFTs the size of their discrete automorphism group. Picking this measure instead will not significantly affect the conclusions reached in this paper.}, and weigh the continuous families by the volume of their conformal manifolds. This gives the following measure over the space of CFTs  in $C(c_{0},\epsilon_c)$
\be \label{measuresec2}
\int_{C(c_{0},\epsilon_c)} d\mu \equiv \frac{1}{\mathcal{N}} \left( \sum_{\mathcal{C}_i}+\sum_{\mathcal{M}_j} \int_{\mathcal{M}_j} d\mu_j^Z \right) \,,
\ee
where the sum over $i$ runs over isolated CFTs, while the sum over $j$ runs over conformal manifolds and $\mu_j^Z$ is the associated measure coming from the Zamolodchikov metric. We note that there can be dualities (like T-duality or S-duality) on a given conformal manifold, that relate theories that a priori one would have thought are different. To only average over distinct CFTs, the conformal manifold should thus always be quotiented by the action of the duality group. The normalization factor $\mathcal{N}$ assures that the measure integrates to one and reads
\be
\mathcal{N} \equiv  \left( \sum_{\mathcal{C}_i}+\sum_{\mathcal{M}_j} V_j \right) \,,
\ee
where $V_j$ is the volume of the conformal manifold with respect to the Zamolodchikov metric. Here, we give no input on the CFTs other than the Zamolodchikov metric, which is arguably the most natural metric on a conformal manifold. This probability measure can thus be viewed as the maximum ignorance ensemble on $C(c_{0},\epsilon_c)$.

This measure is meaningful if the normalization factor $\mathcal{N}$ is finite, i.e. if the measure is normalizable. Unfortunately, it is not finite across all values of the central charge and at high central charge, it will generally be divergent. This will force us to regularize the measure, and as we will see, a natural way to do so is be introducing a lower-bound on the spectral gap. But first, we test our measure in a context where there exists a complete classifications of CFTs: CFTs with $c\leq 1$.

\subsection{CFTs with $c\leq1$}
We would like to understand the space of CFTs for $c\leq 1$. For example, we can consider $c_0=3/4$ and $\epsilon=1/4$. Thankfully, CFTs with $c<1$ are fully classified: they are the Virasoro minimal models (see for example \cite{DiFrancesco:1997nk}). At $c=1$, CFTs are also classified: there is the compact free boson and its $\mathbb{Z}_2$ orbifold, along with 3 isolated CFTs obtained from orbifolding the $SU(2)$ WZW theory at level 1. The free boson is not an isolated theory, as the compactification radius is arbitrary. In fact, the compactification radius is a modulus of the theory, and one can change radius by deforming the CFT by the exactly marginal $:J \bar{J}:$, where $J$ is the $U(1)$ current of the theory.

Since the space of CFTs below $c=1$ is well understood, we can try and define the measure $\mu$ over all such theories, namely the space $C(c_0=3/4,\epsilon=1/4)$. As we will now see, the problem is that this measure is poorly defined, since $\mathcal{N}$ diverges. Both its isolated (i.e. the minimal models) and continuous parts (i.e. the free boson) diverge. For the minimal model, this is because there are infinitely many minimal models, with an accumulation point at $c=1$. For the free boson, it is because the volume of moduli space using the Zamolodchikov metric diverges for a single free boson. Both these divergences can be regulated by introducing a lower bound on the spectral gap, as we will now see explicitly. We will start with the minimal models.

\subsubsection*{The Virasoro minimal models}

We start by quickly reviewing some features of the Virasoro minimal models, see \cite{DiFrancesco:1997nk} for more details. These theories are unitary CFTs with
\be
c=1-\frac{6}{m(m+1)} \,,
\ee
for any $m\geq3$. They have a finite number ($m(m-1)/2$) of Virasoro primaries, whose weights are
\be
h_{r,s}=\frac{((m+1)r-m s)^2-1}{4m(m+1)} \,,
\ee
with the integers $1\leq r\leq m$ and $1\leq s\leq m+1$. There are infinitely many minimal models which accumulate to $c=1$ as $m\to\infty$.\footnote{There are in fact several families of minimal models, depending on how left and right moving representations are combined. For simplicity, we will only keep track of the diagonal minimal models which have only scalar operators. Adding the other families will not substantially change the properties described below.} We will soon be interested in regulating the number of CFTs, and a natural way to proceed is to impose a minimal gap.

It is easy to see that for the minimal models, the lightest operator is $h_{2,2}$. It has dimension
\be
\Delta_{\min}=2h_{2,2}=\frac{3}{2m(m+1)} \,.
\ee
The number of CFTs with a gap no smaller than a given $\Delta_{\min}$ is thus
\be
\mathcal{N}_{\textbf{MM}}\sim \Delta_{\min}^{-1/2} \,.
\ee
At fixed $\Delta_{\min}$, we thus have a finite number of CFTs. We now turn to the free boson.

\subsubsection*{The compact free boson}

The partition function of the free compact boson is given by
\be \label{freebosonZ}
Z(\tau,\bar{\tau})=\frac{1}{|\eta(\tau)|^2}\sum_{e,m \in \mathbb{Z}} q^{(e/R+mR/2)^2/2}\bar{q}^{(e/R-mR/2)^2/2} \,,
\ee
where $q=e{2\pi i \tau}$ and the boson is compactified such that $\phi\sim \phi+2\pi R$. The integers $e,m$ label electric/magnetic charges and the weights and spin read
\be \label{spectrumfreeboson}
\Delta_{e,m}=e^2/R^2+\frac{m^2R^2}{4} \,, \qquad J_{e,m}=e m \,.
\ee
The free boson admits an exactly marginal deformation, induced by the operator $:J\bar{J}:$ where $J$ is the $U(1)$ current. The effect of the deformation is to change the radius of the boson. The conformal manifold of the theory is given by the different choices of radii $R$, and the Zamolodchikov metric is up to order one factors
\be
ds_Z^2\approx \frac{dR^2}{R^2} \,.
\ee
The free boson also has a T-duality symmetry, and is left invariant under the action
\be
R\to\frac{2}{R} \,,
\ee
as can explicitly be seen in \rref{freebosonZ}. To count only distinct CFTs, one should therefore restrict the range of $R$, and without loss of generality, we will pick the range
\be
R\in [\sqrt{2},\infty) \,.
\ee
The volume of the conformal manifold diverges due to the decompactification limit $R\to\infty$. Introducing a maximal radius $R_{\max}$, we have
\be
V_{\text{free boson}}=\int_{\sqrt{2}}^{R_{\max}} \frac{dR}{R} \sim \log R_{\max} \,.
\ee
Using the relation between the spectrum and the radius \rref{spectrumfreeboson}, we have $R_{\max} \sim \Delta_{\min}^{-1/2}$ and thus
\be
V_{\text{free boson}} \sim - \log \Delta_{\min} \,.
\ee
The divergence is slower than that coming from the minimal models. This means that the minimal models dominate the space of CFTs in the ensemble below $c=1$.\footnote{It is interesting to try and interpret the limit of the minimal models as a genuine CFT at $c=1$, which would have a continuous spectrum. See \cite{Runkel:2001ng} for a discussion of this, or \cite{Fredenhagen:2012rb,Fredenhagen:2012bw} for the supersymmetric case.} Since the measure is normalizable at fixed $\Delta_{\min}$, an average over theories can be performed. This was studied for example in \cite{Benjamin:2018kre}, in particular the spectral form factor of this ensemble. We now discuss the space of CFTs at higher values of $c$.

\subsection{The landscape at higher $c$}

For $c>1$, it is much harder to make definitive statements as the landscape of CFTs is still poorly understood. Nevertheless, the results we have already obtained, combined with the fact that we can build new CFTs at $c>1$ by tensoring together CFTs with $c\leq 1$ already teach us a lot about the landscape. For example, we already know that there will be an accumulation point of infinitely many CFTs at $c=c_{\text{known}}+1$, by considering the CFTs
\be
\mathcal{C}= \mathcal{C}_{\text{known} }\otimes \text{MM}_{m} \,, \qquad \text{or} \qquad \mathcal{C}_{\text{known}} \otimes \text{FB}_R \,,
\ee
where MM$_m$ and FB$_R$ are the $m$-th minimal model and free boson at radius $R$, respectively. Note that the divergence in the number of CFTs may not always come from both the free boson construction and the minimal models construction. This is because of the fact that if $\mathcal{C}_{\text{known}}$ contains itself a free boson CFT in it, the conformal manifold of the tensor product is enlarged, and in particular has finite volume. Already for two free bosons, the conformal manifold has a finite volume \cite{Borel1962}.

It trivially follows that there are accumulation points of CFTs at all integer $n$, by considering
\be
\mathcal{C}= \text{FB}^{\otimes n-1} \otimes \text{MM}_m \,,
\ee
and at half integer $n+1/2$ with $n\geq 1$ by tensoring-in an additional copy of the Ising model
\be
\mathcal{C}=\text{MM}_3 \otimes \text{FB}^{\otimes n-1} \otimes \text{MM}_m \,.
\ee

In fact, simply by tensoring in multiple copies of the minimal models, one can show that the accumulation points of CFTs become extremely dense. In fact, this leads us to our first result on the space of CFTs (see Appendix \ref{app:accumulationpoints} for the derivation).

\begin{result}
The central charge gap between any two accumulation points of CFTs can be no bigger than 
\be \label{CFTgap}
\delta c_{\text{accum}}\leq\left(\frac{3}{2}\right)^{-\left(3/2\right)^{c}} \,, \quad c\to\infty
\ee 

\end{result}

Note that this is doubly non-perturbative in the central charge! This result is about the density of accumulation point of CFTs, so it means that the difference in central charge between two values of $c$ at which there are infinite number of CFTs is bounded above by something extremely small. This means that the space of CFTs is plagued by a very dense set of points where our naive measure diverges. This leads us to a regulation of the measure. 

\subsection{Regulating the measure}

We have seen that already for CFTs close to $c=1$, and close to an extremely dense set of points at larger $c$, the number of CFTs blows up, leading to a divergence of our measure. To regulate this divergence, we introduce a new parameter in our space of CFTs: $\Delta_{\min}$. This is a gap on the spectrum of local operators, meaning we only consider CFTs where all local operators have dimension
\be
\Delta_O \geq \Delta_{\min} \,, \qquad \forall \  O \,.
\ee
This defines for us a new space of CFTs $C(c_0,\epsilon_c,\Delta_{\min)}$ and we have already seen that this measure is normalizable for any $\Delta_{\min}>0$, and $c_0=3/4$, $\epsilon=1/4$ or for any other neighbourhood of CFTs with $c<1$.

In fact, it has been conjectured that the number of CFTs below a given central charge, with fixed $\Delta_{\min}$ is finite \cite{Kontsevich:2000yf,Acharya:2006zw}. The conjecture relies on geometric observations on the moduli space of supersymmetric CFTs corresponding to Calabi-Yau sigma models, and is similar in spirit to other conjectures of the Swampland program \cite{Ooguri:2006in}. For now, we will proceed under the assumption that the conjecture is true, and discuss the alternative when we come to 3D gravity below. Before turning to 3D gravity, we will mention other results and properties for the landscape of CFTs.

\subsection{Other results about the landscape of CFTs}

Above, we have found accumulation points of CFTs: values of the central charge where an infinite number of CFTs live. So far, all the CFTs we have discussed are rational. A natural question to ask is whether there are accumulation points of irrational CFTs. What we will now show, is that there are accumulating conformal manifolds on which irrational points exist.

In general, the space of non-rational CFTs is sadly not well understood. For example, one does not even know of a concrete CFT with $c>1$ and only Virasoro symmetry.\footnote{See however \cite{Antunes:2022vtb,Kousvos:2024dlz,Antunes:2024mfb,Antunes:2025huk,Antunes:2025erb}, based on ideas from \cite{Dotsenko:1998gyp} for progress on obtaining such a CFT through a relevant deformation or products of minimal models.} The non-rational CFTs that are best understood are supersymmetric, the most famous of which is the D1D5 CFT (see \cite{David:2002wn} for a review). These theories are obtained by deforming a symmetric product orbifold (we will give a review of symmetric orbifolds in the following section) of $T^4$ or K3, by an exactly marginal operator. The marginal operator is guaranteed to be exactly marginal by supersymmetry, and gives anomalous dimensions to the higher spin currents in perturbation theory, rendering the theories irrational \cite{Gaberdiel:2015uca,Apolo:2022fya}. Beyond the D1D5 CFT, we also know of infinitely many other examples, by taking symmetric product orbifolds of the $\mathcal{N}=(2,2)$ minimal models or their tensor products \cite{Belin:2020nmp,Apolo:2022fya,Benjamin:2022jin} and applying similar deformations. 

Unfortunately, these theories are not well understood beyond their study in conformal perturbation theory starting from the orbifold point, see \cite{Gaberdiel:2015uca,Hampton:2018ygz,Keller:2019suk,Guo:2019pzk,Guo:2019ady,Guo:2020gxm,Benjamin:2021zkn,Apolo:2022fya,Benjamin:2022jin,Keller:2023ssv}. For example, the global structure of the conformal manifolds are poorly understood. Nevertheless, the mere existence of such theories already teaches us interesting things about the landscape of non-rational CFTs. Using the construction of \cite{Belin:2020nmp}, we obtain our second result about the landscape of CFTs:

\begin{result}

There are non-trivial accumulation points of irrational CFTs. This simplest example appears at $c=9$. This accumulation point is described by taking the CFTs
\be \label{symcube}
\frac{\text{SMM}_{k} \otimes \text{SMM}_{k}   \otimes \text{SMM}_{k} }{S_3} \,, \quad k\to\infty 
\ee
and deforming away from the orbifold point by a twist-3 operator.
\end{result}
Here, $k$ is a label for the supersymmetric minimal models, similarly to $m$ for the bosonic ones. The deformation operator that turns the orbifold theory into an irrational CFT exists at any $k$, and these CFTs have central charge
\be
c=\frac{9k}{k+2} \,,
\ee
thus the central charge asymptotes to $9$ in the large $k$ limit. Note that here, we really find an accumulation point of conformal manifolds, and our claim is that it contains irrational 2d SCFTs. 

Of course, each of these theories has a very interesting (and complicated) conformal manifold, and we cannot really talk about an accumulation point of non-rational CFTs per say, since one should average over the conformal manifold of such theories, and the notion of rationality will not be invariant on the conformal manifold. Nevertheless, this establishes that there is an accumulation point of CFTs near $c=9$ that have conformal manifolds where some (and probably most) points are irrational.

Another aspect worth noting is that these theories have conformal manifolds of very large dimension, \cite{Belin:2020nmp}. In fact, this leads us to our third result on the landscape
\begin{result}
There exists conformal manifolds of CFTs of bounded central charge, where the dimensionality of the conformal manifold is unbounded.
\end{result}
For example, for the theories \eqref{symcube} the dimension of the conformal manifold scales like $k^2$. The question of bounding the number of marginal operators was discussed from a modular bootstrap perspective in \cite{Hellerman:2010qd}, where they argued that it would be surprising to find theories where the number of marginal operators is arbitrarily large while the central charge is bounded. Naturally, these theories do not violate any of the modular bootstrap bounds, and the loophole comes from the large number of very relevant operators, whose dimensions approach zero in the large $k$ limit.

Note that these theories could also give a problem for the normalization of the measure, depending on how the volume of the conformal manifolds scale in this limit. Fortunately, introducing a spectral gap $\Delta_{\min}$ cures both problems. The properties of the spectrum of the $\mathcal{N}=(2,2)$ minimal models follows closely that of their bosonic cousins, and there are BPS operators (hence protected on the conformal manifold) that have very small scaling dimensions as $k$ becomes large. Therefore, putting a minimal spectral gap will first cap off the number of distinct conformal manifolds. In doing so, it also caps off the theories whose conformal manifold has unbounded dimension.

An interesting question is whether we can make any statement about the gap in central charge between irrational CFTs. What is the upper bound we can currently construct? The best that we can do, is take the $N$-fold symmetric orbifold of the SMM$_1$, which has a unit gap between theories. At any $N>5$, the theory has an exactly marginal operator that can make the theory irrational \cite{Belin:2020nmp}. Therefore, we have a gap of size one in central charge between two irrational theories (the one at $N$ and the next one at $N+1$). We can also take the symmetric orbifold of other minimal models, or products of minimal models. This will decrease the gap, but still make it finite in the large central charge limit. It would be interesting to see if we can make it smaller.

\section{Implications for pure gravity in AdS$_3$ \label{sec:gravinter}}

We now discuss the implications for pure 3D gravity in AdS$_3$. The hope is that pure 3D gravity can be obtained by integrating over our probability measure, namely
\be
Z_{AdS_3} = \frac{\int_{C(c_0,\epsilon_c,\Delta_{\min})} d\mu  Z_{CFT}}{\int_{C(c_0,\epsilon_c,\Delta_{\min})} d\mu} \,,
\ee
with
\be
c_0 = \frac{3 \ell_{AdS}}{2G_N} \,.
\ee
This would mean that pure 3D gravity is the maximum ignorance ensemble over all CFTs, and that a typical draw from our ensemble would have the properties of 3D gravity, like a gap of order $c_0$ to the first Virasoro primary.

The first issue, is that on the quantum gravity side, there is no notion of the parameters $\epsilon_c$ and $\Delta_{\min}$. In the end, these parameters were both regulators of some type and it is natural to think that the bulk theory must correspond to some limit where these parameters disappear.  We will start by discussing the limit $\Delta_{\min}\to 0$. We would thus like to understand what happens in the limit 
\be
\lim_{\Delta_{\min}\to 0}\frac{\int_{C(c_0,\epsilon_c,\Delta_{\min})} d\mu  Z_{CFT}}{\int_{C(c_0,\epsilon_c,\Delta_{\min})} d\mu} \,.
\ee

To understand the outcome, we first go back to conjecture of Kontsevich and Soibelman, which states that the space of CFTs is finite as long as $\Delta_{\min}$ remains finite. Let us start by assuming that this conjecture is true. If this is indeed the case, it means in particular that there are finite number of CFTs that look like pure gravity: large gap in operator dimensions, no extra currents, etc. Now at some fixed $\Delta_{\min}$, it may well be the case that the number of pure gravity like theories highly dominates over any other type of CFT (for example the product of minimal models). But as we decrease $\Delta_{\min}$, this number stays constant while the number of products of minimal models start to drastically increase. At some (potentially much smaller) value of $\Delta_{\min}$, the products of minimal models now dominate the average. In any case, in the limit
\be
\lim_{\Delta_{\min}\to 0}\frac{\int_{C(c_0,\epsilon_c,\Delta_{\min})} d\mu  Z_{CFT}}{\int_{C(c_0,\epsilon_c,\Delta_{\min})} d\mu} \,,
\ee
we are guaranteed that the ensemble will be dominated by the product of minimal models, and that true pure gravity-like theories will be completely washed away by the proliferation of the other theories. In particular, in the average, things like the spectral gap will be small (we expect it to vanish in this average). It is therefore not possible to obtain 3D gravity in a limit of this kind.

Another possibility is that the conjecture of Kontsevich and Soibelman is wrong, and that there are an infinite number of CFTs, even at fixed $\Delta_{\min}$. If this is the case, then we have a more serious problem: we have failed to define a normalizable probability measure. It is still possible that there could be another way to regulate the measure such that we can define a mathematically meanfingfull average. We discuss several other options in the discussion section.

Given the difficulties encountered already with the $\Delta_{\min}\to0$ limit at fixed $\epsilon_c$, the fate of the $\epsilon_c\to0$ is perhaps less relevant. We expect that taking this limit will yield a very erratic probability measure at fixed $c_0$, since it will start to probe the quantization of the cosmological constant. If we keep $\epsilon_c$ fixed, then we have some type of spread or variance for the central charge. A scenario like this has been discussed in \cite{Schlenker:2022dyo}, but we are not aware of any bulk computation that would probe the variance of the cosmological constant, or that would establish the finiteness of $\epsilon_c$ from the bulk perspective.\footnote{It is worth noting that for string theory on AdS$_3\times S^3 \times T^4$, the bulk theory is proposed to be at fixed chemical potential for the central charge, rather than at fixed central charge (see for example \cite{Eberhardt:2021jvj}).}

To conclude, we have found that, at least for the probability measure we defined, it is simply not true that pure gravity in AdS$_3$ is the average over the maximum ignorance ensemble of CFTs. A typical draw from that ensemble is in fact very different from pure gravity, and has a huge proliferation of light operators. This means that pure gravity theories (if they exist) are somehow highly atypical within the space of all CFTs. 

One may wonder whether a more general statement is true, that holographic CFTs (not necessarily pure gravity-like, but theories dual to semi-classical GR with minimally coupled matter) are also atypical in the space of all CFTs. To test this idea in settings where the landscape of CFTs is well understood, we now discuss its realization for one particular landscape of CFTs: permutation orbifolds.

\section{Permutation orbifolds \label{sec:permutation}}

We would now like to explore the ideas discussed in this paper in concrete families of CFTs that are under control, where quantitative statements can be made. It is in general quite difficult to build landscapes of CFTs for which the holographic properties of each single theory can be well tracked. However, permutations
orbifolds have the useful property that many of their holographic characteristics are governed by group theory. When starting with a seed CFT
$\mathcal{C}$, and taking the tensor product of $N$ copies of it $\mathcal{C}^{\otimes N}$, the resulting
theory has a huge global symmetry coming from all the ways we can permute the $N$ copies. This symmetry is given by the symmetric group $S_{N}$. One can then gauge the resulting $S_{N}$ global symmetry (or any subgroup of it), and obtain
the permutation orbioflds 
\[
\mathcal{C}_{N}\equiv\frac{\mathcal{C}^{\otimes N}}{G_{N}},\quad\quad G_{N}\subseteq S_{N}.
\]
From just a single seed CFT $\mathcal{C}$, we have produced a huge landscape of new CFTs, where each new CFT is associated to a choice of subgroup $G_{N}$.
By taking $N$ large, we can study these theories in the regime which would correspond to the semiclassical limit in a putative bulk dual
\[
c_{N}=Nc_{\text{seed}}\,.
\]

These theories are an interesting case study in light of holography.
Indeed, many of their properties at large $N$ are insensitive to the
choice of seed CFT, but instead depend entirely on the choice of subgroup
$G_{N}$. For example, one expects a
holographic CFT to have have a finite number of states at any fixed scaling dimension $\Delta$ in the large $N$ limit. For permutation orbifolds, this condition is purely group theoretic
and reduces to demanding that $G_{N}$ has a finite number of orbits on $k$-tuples
in the large $N$ limit \cite{Belin:2014fna,Haehl:2014yla}. It turns
out that this property has been studied by group theorists for subgroups
of $\text{Sym}(\mathbb{N})$. Subgroups of this type are called oligomorphic
on $\mathbb{N}$ \cite{Cameron1990}. The notion of oligomorphic families
\cite{Belin:2015hwa} has been proposed to capture the notion of families
$\{G_{N}\}$ of finite subgroups of $S_{N}$ such that 
\[
\underset{N\rightarrow\infty}{\lim}G_{N}=G
\]
where $G$ is an oligomorphic group on $\mathbb{N}$.

Another property which is typically assumed of holographic CFTs is the uniqueness of the stress-tensor, namely that the stress-tensor is the unique quasi-primary operator of spin-2. For permutation orbifolds, whether this condition is met or not depends again on the underlying group structure: the group
$G_{N}$ must have a unique orbit on $1$-tuples, i.e $G_{N}$ is
a transitive subgroup of $S_{N}$.

\vspace{1em}

It is important to note that subgroup transitivity is an independent condition from oligomorphicity: there are oligomorphic groups that are not transitive (for example $S_{N/2}\times S_{N/2}\subset S_N$), and there are transitive groups that are not oligomorphic (like $\mathbb{Z}_N$). For our purposes, we will demand both, and see that in fact transitivity alone is enough to already derive interesting bounds.

The set of subgroups $G_{N}\subseteq S_{N}$ whose permutation
orbifolds $\mathcal{C}_{N}$ we will be interested in is thus
\[
O_{N}=\{\text{groups which are oligomorphic as }N\rightarrow\infty\}\bigcap\{\text{transitive subgroups of }S_{N}\}.
\]

The question ``Within the landscape of permutation orbifolds, how many of them have holographic properties?"  can thus be restated as a subgroup counting problem: What fraction of subgroups of $S_N$ does $O_N$ represent? If this fraction vanishes in the large $N$ limit (as we will see it does), we are led to conclude that holographic-like permutation orbifolds are rare in the large $N$ limit.

\subsection{Subgroup number and conjugacy classes}

It is useful to mention some preliminaries about what exactly we mean by counting subgroups here, and some known results from the mathematical litterature.
The number of subgroups of $S_{N}$, denoted $s(S_{N})$, traditionally counts all possible ways to construct a subgroup of $S_{N}$ using its elements. Therefore, if the group $G_{N}$ has several different embeddings in $S_{N}$, $s(S_{N})$ will count each of these as different subgroups. To date, the best known bounds on $s(S_{N})$ are found in \cite[Corollary 3.3 and Theorem 4.2 respectively]{Pyber1993,Pyber1993a},\footnote{We are using the little $o$ notation, where $f(x)=o(g(x))$ is defined as $\left|f(x)\right|\leq\epsilon g(x)$ for any $\epsilon>0$ as $x\rightarrow\infty$. In other words, $f(x)=o(g(x))$ effectively means that $f$ grows much slower than $g$.}
\begin{equation}
\boxed{2^{\left(1/16+o(1)\right)N^{2}}\leq s(S_{N})\leq24^{\left(1/6+o(1)\right)N^{2}}},\label{eq:SN bound}
\end{equation}
In fact, it is even
conjectured\cite{Pyber1993a,Pyber1993} that 
\begin{equation}
s(S_{N})=2^{\left(1/16+o(1)\right)N^{2}}.\label{eq:conjecture Sn subgroup number}
\end{equation}

However, this is the total number of subgroups, which discriminates among different embeddings of a same group $G_{N}$. Some of these might have similar or different numbers of orbits when acting on $k$-tuples. If we look at the groups $\{1,(12)\},\{1,(13)\}$ and $\{1,(12)(34)...(N-1\,N)\}$, these are all different embeddings of the cyclic group $\mathbb{Z}_{2}$. When acting on the 1-tuples $\{i\},i=1,...,N$, we can nevertheless see that both $\{1,(12)\}$ and $\{1,(13)\}$ have $N-1$ orbits, while $\{1,(12)(34)...(N-1\,N)\}$ has $N/2$ orbits. Therefore,
when studying the set of subgroups of $S_{N}$ for orbifold purposes, what we are actually interested
in is the number of subgroups with similar group action. Such subgroups
(in their permutation group embedding) are called permutation isomorphic,
which has been shown to be equivalent to the subgroup embeddings being conjugate
\cite{Dixon2012}, and we are thus interested in the number of subgroups
up to conjugacy, which we will denote $c(S_N)$. As mentioned in \cite[Section 6]{Pyber1993a}, and
by noting that there can be at most $N!$ subgroups in a same conjugacy
class, we have similar asymptotics as \ref{eq:SN bound} for the number
of subgroup conjugacy classes, as we have 
\begin{align}
\frac{s(S_{N})}{N!} & \leq c(S_{N})\leq s(S_{N})\nonumber \\
\Rightarrow2^{\left(1/16+o(1)\right)N^{2}-N\log N} & \leq c(S_{N})\leq24^{\left(1/6+o(1)\right)N^{2}}\nonumber \\
\Rightarrow2^{\left(1/16+o(1)\right)N^{2}} & \leq c(S_{N})\leq24^{\left(1/6+o(1)\right)N^{2}} \,,
\end{align}
asymptotically. As we will see shortly, there are results about the densities of transitive
and oligomorphic subgroups when considering
them up to conjugacy.

\subsection{The transitivity and oligomorphy restrictions}

In order to estimate the density of $O_{N}$, we first look at the
transitivity requirement. It has been shown in \cite{Lucchini2000} that,
for $s_{\text{trans}}(S_{N})$ the number of transitive subgroups
of $S_{N}$, we have the upper bound
\begin{equation}
s_{\text{trans}}(S_{N})\leq2^{b\frac{N^{2}}{\sqrt{\log N}}},\label{eq:transitivity bound}
\end{equation}
for any $N$ and some constant $b$.\footnote{The constant has not been determined (nor has it been numerically bounded, to the best of our knowledge). The theorem proves that such a constant exists.} We can extend this result to
conjugacy classes of transitive subgroups, since $N^{2}/\sqrt{\log N}\gg N\log N$.
From this, we see that the density of transitive subgroups (up to
conjugacy) goes to 0.
\be
\frac{c_{\text{trans}}(S_{N})}{c({S_N})} \sim  e^{\frac{N^2}{\sqrt{\log N}}-N^2} \sim e^{-N^2} \,, \qquad N\to\infty \,.
\ee

This already proves that holographic-like permutation orbifolds
are sparse within the set of all orbifold theories. Nevertheless, the corrections to the leading asymtpotic are large, and given by $\sim e^{N^{2}/\sqrt{\log N}}$. What we would like to know is how the oligomorphic condition further constrains the number of subgroups. Unfortunately, this is a difficult question. At the level of oligomorphic groups (i.e. at $N=\infty$) it is still poorly understood which groups come from oligomorphic families, and can therefore
be used to construct permutation orbifolds for finite $N$. This makes the question of density of oligormorphic families as one approaches infinite $N$ very delicate. 

While we will not be able to derive an upper bound on the number of oligomorphic groups, we will derive a lower bound simply by constructing large classes of oligomorphic (and transitive) families. We do this in App. \ref{app:lowerbound}, finding that for any constant $l$ fixed in the large $N$ limit 
\be
c_{\text{oligo}\bigcap \text{trans}}(S_N) \geq (\log N)^{l-1} \,.
\ee
This gives us the estimate
\be
 (\log N)^{l-1}\leq c_{\text{oligo}\bigcap \text{trans}}(S_N) \leq 2^{b \frac{N^2}{\log N}} \,,
\ee
for any fixed $l$ in the large $N$ limit. We see that there is a huge gap between the lower and upper bounds, and it is likely that the number of oligomorphic families is quite a bit larger than what we have been able to construct.\footnote{In appendix \ref{app:lowerbound}, we show that the growth of oligomorphic but non transitive groups is exponentially larger, which is already an important sign in this direction.} At the same time, we do not expect it to come close to the upper bound. One can reasonably expect that holographic permutation orbifolds are extremely sparse within the entire landscape.

Finally, we note that there is also a discussion of ``typical'' subgroups of $S_{N}$ which can be found in the subgroup theory literature, even if only at the state of a conjecture. From \cite[Sections 4 and 6]{Pyber1993a},
the ``typical'' subgroups of $S_{N}$
are conjectured to be elementary abelian 2-groups, which have the form $(\mathbb{Z}_{2})^{q}$
for $q\in\{1,2,...,2/n\}$.There is a large number of these subgroups, and in fact this large number of elementary abelian 2-groups gives rise to the lower bound of \ref{eq:SN bound}. If these groups really saturate the space of subgroups of $S_{N}$, then a typical subgroup has $\sim N/2$ orbits on 1-tuples. This would immediately imply that most subgroups are not oligomorphic, and in fact the density of
oligomorphic subgroups of $S_{N}$ would therefore also decay as $e^{-N^{2}}$ as $N\rightarrow\infty$, as expected.

\section{Discussion \label{sec:discussion}}

In this paper, we have introduced a probability measure over the space of CFTs. It is defined by three parameters: a mean central charge $c_0$, a central charge spread $\epsilon_c$ and a minimum gap $\Delta_{\min}$ for the spectrum of local operators. This measure is normalizable, provided that a conjecture due to Kontsevich and Soibelman is true. The measure we have defined treats all isolated CFTs equally and weighs conformal manifolds locally by their Zamolodchikov metric. We interpret it as the maximum ignorance ensemble over CFTs.

In the process of describing our measure, we have provided several new results on the space of CFTs: a bound on the maximal spacing in central charge between CFTs, results on accumulation points of irrational CFTs and on the dimensionality of conformal manifolds of 2d SCFTs.

We then tested whether pure gravity could be obtained as the average over all CFTs using our measure. Unfortunately, we saw this was not the case, and either the conjecture of Kontsevich and Soibelman is true and a typical CFT does not look like pure gravity, or the conjecture is false and we do not know how to produce a normalizable measure. This implies that pure gravity-like theories are highly atypical within the space of CFTs, which resonates with our results on permutation orbifolds. For those theories, we found that permutation orbifolds with a single stress-tensor and a finite number of operators of fixed dimension in the large $N$ limit are highly atypical.

We conclude with some workarounds and open questions.

\subsection{Twist gap}

Another way to regulate the measure is to introduce a twist gap in the spectrum of Virasoro primaries. This means we want CFTs for which all primary operators have\footnote{Here, we are not talking about the value of the gap in the large spin limit, but really as a statement at finite spin.}
\be
\min(h,\bar{h}) > h_{\min} \,.
\ee
This also implies a bound on scaling dimensions, but is a much stronger condition. It implies that the theory has no extra currents and that the chiral algebra is only Virasoro. It is possible that for a regulator of this type, it becomes true that a typical CFT is pure gravity like: all the products of minimal models get ruled out since taking the tensor product of two CFTs always generates new currents. 

This is an interesting possibility worth exploring. Unfortunately, with our current understanding of the space of CFTs, we cannot make any quantitative statements. The space of CFTs with a twist gap is extremely poorly understood, and up to the construction of \cite{Antunes:2022vtb}, we did not even know of a single example. 

A second reason why this does not appear appealing to us, is that this is imposing by hand some strong property in the CFT. This is far from a maximum ignorance ensemble over all CFTs, since this restricts the chiral algebra. To make an analogy with random matrix theory, there we do not impose by hand that we exclude integrable quantum systems, they are just highly disfavored by the matrix integral. For example, the probability to have degenerate eigenvalues is measure zero. But there are entire regions of almost degenerate levels that have finite measure, even if they don't contribute much to the probability distribution. If the parallel with chaotic Hamiltonian goes through, it seems natural to expect something similar here, where rational CFTs with extra currents should be disfavored by the probability measure, not excluded by hand.

\subsection{Chemical potentials for $\Delta_{\min}$ and $\epsilon_c$}

We have regularized the measure by introducing a window in central charge to avoid erratic jumps as we change $c_0$, and a gap in scaling dimensions to make the measure normalizable. Another way to accomplish the same effect, is to introduce chemical potentials for both. We could try to introduce a chemical potential for the central charge, similar to how the DMVV formula works in symmetric orbifolds \cite{Dijkgraaf:1996xw}. This would fix the expectation value of the central charge in the ensemble, rather than a strict central charge window. One issue is that the number of CFTs grows very quickly with central charge, so the chemical potential-like term could not be a simple exponential like in statistical mechanics.

Similarly, one can introduce a chemical potential for the gap size. We would add to the probability measure a factor
\be
e^{-\frac{\mu_{\text{gap}}}{\Delta_{\min}} } \,.
\ee
This would allow to always average over the full space of CFTs $C(c_0,\epsilon_c)$, and the divergence at small gap would be killed by the exponential term. We still expect similar issues for the purposes of 3D gravity, as the regulator $\mu_{\text{gap}}$ would then be sent to zero, but it would be interesting to see how the average behaves as a function of it.

\subsection{Double scaling for gravity?}

We now comment on the possible ways to still have 3D gravity be the average over all CFTs. One possibility, is of course to have a fixed $\Delta_{\min}$ in the ensemble. Then, it is possible that pure gravity-like theories dominate over the space of all CFTs with fixed gap. But it would not be very satisfying to have a parameter on one side of the ensemble, and not on the gravity side. On the gravity side, nothing forbids fields with small mass. As long as they are above the BF bound, AdS is stable and there is no justification for the gap. This makes the gravity side of the ensemble at fixed $\Delta_{\min}$ questionable.

Another possibility which is more interesting, is that gravity emerges in some scaling limit. In this scaling limit, we would scale $\Delta_{\min}\to0$ as $c\to\infty$.\footnote{A natural choice for the size of the gap in this limit is $\Delta_{\min}\sim e^{-c}$ which, provided we can interpret the light fields as free fields in the bulk, would guarantee that the number of bulk states at low energies satisfies the HKS bound \cite{Hartman:2014oaa}.} Then, the parameter disappears in the large $c$ limit, and provided the number of pure gravity-like CFTs grows faster than the rate at which the products of minimal models are polluting the ensemble, we could still get pure gravity in the end. The only objection to this idea, is the fact that pure 3D quantum gravity, on any fixed topology, appears to make sense at finite central charge \cite{Collier:2023fwi,Belin:2023efa}. It is thus puzzling that we would only be able to make a connection in the large $c$ limit. Nevertheless, it would be interesting to pursue this further.

\subsection{Permutation Orbifolds and Oligomorphic families}

We have seen that subgroups of $S_N$ that are transitive and have a well-defined large $N$ limit are highly atypical. It is important to note that the bound we have found, namely that the probability to have an oligomorphic subgroup that is transitive of
\be
p \sim e^{-N^2} \,,
\ee
is only coming from using the transitivity condition. One should also note that while this is the dominant scaling at large $N$, the subleading growth is huge and the total probability is really
\be
p \sim e^{\frac{N^2}{\sqrt{ \log N}}-N^2} \,,
\ee
The term $^{\frac{N^2}{\sqrt{ \log N}}}$ comes from the number of transitive groups, many of which are not oligomorphic. The oligomorphic condition is clearly much stronger, and would reduce this subleading growth drastically. At the level of explicit constructions, we have found a growth which is faster than any power of $\log N$, so essentially a polynomial growth in $N$. This leaves a huge gap in $N$-scaling between what we can construct and the bound coming from transitivity alone.

It would be very interesting to bridge this gap. The issue is that the oligomorphic condition is something that can only be phrased in a limiting procedure as $N\to \infty$. It is very challenging to directly try to impose an oligomorphic condition on subgroups abstractly, as it requires to define sequences in increasing values of $N$. The most hopeful avenue would be to find a criterion that guarantees oligomorphicity, and can be phrased as an expectation value over all subgroups to estimate the probability this way. We hope to return to this question in the future.

%\section{Chiral CFTs and lattices}

%Although making statements about general classes of CFTs is difficult, there is one important class of CFTs -- chiral CFTs -- where the space of CFTs can, at least with a certain assumption, be studied precisely.  In this section we will consider CFTs with $c_L=c\in 24 \mathbb{Z}$ and $c_R=0$.  Our discussion will generalize in a completely straightforward manner to products of chiral and anti-chiral CTs.

%Let us consider a chiral CFT with central charge $c=24k$, with $k\in \mathbb{Z}$.  We will assume that the theory is compact, modular invariant, and has a unique ground state.  Since the leftmoving dimensions vanish (${\bar h}=0$) all states have integer dimension $h\in \Z$.  Uniqueness of the vacuum therefore implies that all non-vacuum states have $h\ge 1$.  So chiral CFTs automatically have a gap in the spectrum of conformal dimensions.  Based on our earlier conjecture, we would expect that the number of chiral CFTs is finite.  We can now proceed to check our conjecture in this case.

%To begin, we note that there is one simple family of chiral CFTs, those constructed from $c$ free chiral bosons which live on $\mathbb{R^c}/\Lambda$ where $\Lambda$ is an even self-dual lattice of signature $(c,0)$.  }

\section*{Acknowledgements}

We are happy to thank Connor Behan, Nathan Benjamin,  Suzanne Bintanja, Jose Calderon-Infante, Scott Collier, Jan de Boer, Tolya Dymarsky, Lorentz Eberhardt, Victor Gorbenko, Thomas Hartman, Christoph Keller, Diego Liska, Miguel Montero, Kyriakos Papadodimas, Eric Perlmutter, Boris Post, Julian Sonner and Irene Valenzuela for stimulating discussions. The work of FS was supported by the Laboratory for Theoretical Fundamental Physics (LTFP) at the Ecole Polytechnique Federale de Lausanne, as well as the research grant 00025445 from Villum Fonden and by the Sapere Aude: DFF-Starting Grant 4251-00029B. This work was made possible by Institut Pascal at Université Paris-Saclay with the support of the program “Investissements d’avenir” ANR-11-IDEX-0003-01. 
Research of AM is supported in part by the Simons Foundation Grant No. 12574 and the Natural Sciences and Engineering Research Council of Canada (NSERC), funding reference number SAPIN/00047-2020

\appendix

\section{Distribution of Accumulation points \label{app:accumulationpoints}}
In this appendix, we bound the density of accumulation points of 2d CFTs.
Given the $m$-th minimal model $\text{MM}_{m}$ with central charge
\begin{equation}
c_{m}=1-\frac{6}{m(m+1)},\,m\geq3
\end{equation}
where we find an accumulation of points towards $c_{\infty}=1$ as
$m\rightarrow\infty$, we can construct infinitely many accumulation
points by taking the tensor product of $n$ minimal models with fixed
$m_{i}$, $i\in\{1,...,n\}$ with a tensor product of $k$ accumulation
sequences $\text{MM}_{l_{j}}$, $j\in\{1,...,k\}$, giving a family
of CFTs of the form
\begin{align}
\mathcal{C}_{m} & =\text{MM}_{m_{1}}\otimes...\otimes\text{MM}_{m_{n}}\otimes\text{MM}_{l_{1}}\otimes...\otimes\text{MM}_{l_{k}},\\
c_{\mathcal{C}_{m}} & =n+k-\left(\frac{6}{m_{1}(m_{1}+1)}+...+\frac{6}{m_{n}(m_{n}+1)}+\frac{6}{l_{1}(l_{1}+1)}+...+\frac{6}{l_{k}(l_{k}+1)}\right).
\end{align}
This gives an accumulation point at 
\begin{equation}
c_{\mathcal{C}}=n+k-\left(\frac{6}{m_{1}(m_{1}+1)}+\dots +\frac{6}{m_{n}(m_{n}+1)}\right).
\end{equation}
\linebreak{}
Note that, using this construction and fixing $k=1$, we can for example
show that if
\begin{equation}
\mathcal{C}'_{m'}=\mathcal{C}_{m}\otimes\text{MM}_{m'},
\end{equation}
then we will get an accumulation point at $c_{\mathcal{C}_{m}}+1$.
We can therefore see that points forming an accumulating family in
the interval $[N,N+1)$, where $N$ is an integer, will give accumulation
points with equal separation between them in the interval $[N+1,N+2)$.
The density of accumulation points will therefore asymptotically increase.
In fact, the construction of new points functions recursively, where
any new points are built out of ``squeezing in'' the accumulation
pattern in-between a preexisting accumulation point and itself plus
one. We can thus try to study the asymptotic evolution of the separation
between accumulation points.

To do this, we first look at how we obtain the accumulation points
in a given interval. Note first that, if we take an accumulation sequence
such as the one described above, the accumulation point is bounded
by 
\begin{align}
n+k-\frac{6n}{3\cdot4} & \leq c_{\mathcal{C}}\leq n+k\nonumber \\
\Rightarrow\frac{n}{2}+k & \leq c_{\mathcal{C}}\leq n+k.
\end{align}
To obtain this, we took the first minimal model $n$ times on the LHS, while we took the $m=\infty$ limit $n$ times on the RHS. Therefore, the only tensor products which can give us an accumulation
point between $[N,N+1)$ are those with up to $n=2N-1$.

To bound the gap between accumulation points, we proceed as follows: If we have an accumulation point
$c$ inside $[N,N+1)$, obtained from the accumulating sequence CFT
$\mathcal{C}$, we can tensor $\mathcal{C}$ with any minimal model
$\text{MM}_{m}$, which will give us 
\begin{equation}
c'=c+1-\frac{6}{m(m+1)} \,,
\end{equation}
with the goal to produce a new accumulation point in the interval $[N+1,N+2)$. We are guaranteed that the resulting theory cannot be outside of $[N,N+2)$, because the central charge of each minimal model is smaller than one, but we are not guaranteed to land in $[N+1,N+2)$. If our original CFT had central charge very close to $N$, and we tensored in a minimal model with too low central charge, we would fail to be in the interval $[N+1,N+2)$. The strategy will be to optimize the gap while making sure we are landing in $[N+1,N+2)$.

Let us start with two accumulation points $C_N$ and $C_N'$ in $[N,N+1)$ with asymptotic central charge $c_N$ and $c_N'$, and we define
\be
\delta c_{N}\equiv c_{N}-c'_{N}>0
\ee
We can now consider the new families
\bea
C_{N+1}^m &=& C_N \otimes \text{MM}_m \\
C_{N+1} &=& C_N \otimes \text{MM}_k \,, \quad  k\to\infty \\
C_{N+1}' &=& C_N' \otimes \text{MM}_k \,, \quad  k\to\infty 
\eea
These have asymptotic central charge
\begin{align}
c_{N+1}^{m} & =c_{N}+1-\frac{6}{m(m+1)}\\
c_{N+1} & =c_{N}+1\\
c'_{N+1} & =c'_{N}+1.
\end{align}

We would now like to find an $m_c$, such that for any $m\geq m_{c}$, $c_{N+1}^{m}>c'_{N+1}$. This guarantees we land in the interval $[N+1,N+2)$. This $m_c$ clearly exists, as for sufficiently large $m$, we are arbitrarily close to $c_{N+1}>c_{N+1}'$. We can now show that the separation between all accumulation points has been reduced in the process. Along the way, we will also bound the optimal $m_c$, which is the minimal value of $m$ such that $\delta c'=0$. First consider
\be
\delta c'\equiv c_{N+1}^{m_{c}}-c'_{N+1}=c_{N}-c'_{N}-\frac{6}{m_{c}(m_{c}+1)}=\delta c_{N}-\frac{6}{m_{c}(m_{c}+1)}
\ee
We want $\delta c'>0$, so we demand

\begin{align}
\frac{6}{m_{c}(m_{c}+1)} & <\delta c_{N}\nonumber \\
\Rightarrow m_{c} & >-\frac{1}{2}+\frac{1}{2}\sqrt{1+\frac{24}{\delta c_{N}}} \,. \label{boundonm}
\end{align}
$m$ must be an integer, so $m_{c}$ will be the smallest integer compatible with \rref{boundonm}. To round up, the worse estimate is that it shifts $m_c$ by one, so we have
\be
m_c < \frac{1}{2}+\frac{1}{2}\sqrt{1+\frac{24}{\delta c_{N}}} \,.
\ee
This gives
hence
\begin{align}
\delta c' & \leq \frac{4 \delta c_N ^{3/2}}{3\sqrt{\delta c_N} + \sqrt{24+\delta c_N}}\nonumber \\
 & \approx \sqrt{\frac{2}{3}}\delta c_{N}^{3/2}+\mathcal{O}(\delta c_{N}^{2}) \,,\label{eq: separation to lower bound}
\end{align}
for small $\delta c_N$. We see that the separation of accumulation points gets reduced under the process we described.

One now needs to worry about the separation between the accumulation point we just found, and the next accumulation point. If we simply took the next one to be $C'_{N+1}$, that separation would not noticeably get reduced. Luckily, an upper bound on the separation between $C^{m_c}_{N+1}$ and the next accumulation point is given by looking at the accumulation point $C^{m_c+1}_{N+1}$. The separation between the two is the separation between two neighboring minimal models, which is
\be
\delta c^{m_c}\equiv c_{N+1}^{m+1}-c_{N+1}^{m}=\frac{6}{m_c(m_c+1)}-\frac{6}{(m_c+1)(m_c+2)}=\frac{12}{m_c(m_c+1)(m_c+2)} \,.
\ee
This gives
\begin{align}
\delta c^{m_{c}} & \leq \frac{4 \delta c_N ^{3/2}}{3\sqrt{\delta c_N} + \sqrt{24+\delta c_N}} \nonumber \\
 & \approx \sqrt{\frac{2}{3}}\delta c_{N}^{3/2} \label{eq:separation between first squeezed points} \,.
\end{align}

Wee see that both separations \ref{eq: separation to lower bound} and \ref{eq:separation between first squeezed points} get reduced under the process we have described. Given any initial separation between two accumulation points, this gives the worse possible separation after one iteration of the process. There are of course many other accumulation points in the interval $[N+1,N+2)$, but we have described the largest ones. 

One can then solve the recursive relation which we obtained, namely
\begin{equation}
\delta c_{N+1}\leq\sqrt{\frac{2}{3}}\delta c_{N}^{3/2}\label{eq:separation recurrence relation} \,,
\end{equation}
with $\delta c_0=1$. We find
\begin{equation}
\delta c_{N}\leq\left(\frac{3}{2}\right)^{1-\left(3/2\right)^{N}}.
\end{equation}
Since $N$ tells us the central charge window we are in, we have $c\sim N$ thus
\begin{equation}
\delta c\leq\left(\frac{3}{2}\right)^{1-\left(3/2\right)^{c}},
\end{equation}
which is doubly exponentially decaying.

\section{Lower-bounding the density of known oligomorphic and transitive families}
\label{app:lowerbound}
In this appendix, we will try to find a lower bound on $c_{\text{oligo}\cap\text{trans}}(S_{N})$.
We will proceed by calculating the number of known permutation group
families which are both oligomorphic and transitive. We can then compare
that with our upper bound on the possible number of such groups. The
main restrictriction on constructing such families as of now is oligomorphicity.
The known oligomorphic and transitive groups relevant to holographic
CFTs are direct products and wreath products of already oligomorphic
groups. We can define these groups from their action on $k$-tuples
of $X_{N}=\{1,...,N\}$. First, set $G_{p}$, $H_{q}$ to be two subgroups
of $S_{p},S_{q}$ respectively, each acting on $\{1,...,p\}$ and
$\{1,...,q\}$ respectively. We set $pq=N$. Then: 
\begin{itemize}
\item The wreath product $G_{p}\wr H_{q}\leq S_{N}$ has elements of the
form $\sigma=(g_{1},g_{2},...g_{q},h)$ with $g_{i}\in G_{p}$ and
$h\in H_{q}$, and size $\left|G_{p}\wr H_{q}\right|=\left|G_{p}\right|^{q}\left|H_{q}\right|$.
Any $k$-tuple of distinct elements is defined as a $p\times q$ matrix
$M_{ij}$ with exactly $k$ non-zero elements. The action is then
given by 
\[
\sigma\cdot M_{ij}=M_{g_{j}(i),h(j)},
\]
i.e we permute the elements of each column independently, and then
permute the columns. 
\item For the direct product $G_{p}\times H_{q}\leq S_{N}$ and with size$\left|G_{p}\times H_{q}\right|=\left|G_{p}\right|\left|H_{q}\right|$,
a typical element is of the form $\sigma=(g,h)\in G_{p}\times H_{q}$
with $g\in G_{p}$ and $h\in H_{q}$. A $k$-tuple is again a $p\times q$
matrix $M_{ij}$ with exactly $k$ non-zero elements, and the action
of $\sigma$ on $M_{ij}$ is then
\[
\sigma\cdot M_{ij}=M_{g(i),h(j)},
\]
i.e we first permute rows, then columns.
\end{itemize}
From \cite{Keller2019,Cameron1990,Cameron2009}, we know that subgroups
of the form $S_{\sqrt{N}}\wr S_{\sqrt{N}}$ and $\bigotimes_{k=1}^{d}S_{N^{1/d}}$
are oligomorphic and transitive if doted with these specific group
actions. In general, from these definitions, we can deduce the conditions
on $G_{p}$ and $H_{q}$ such that $G_{p}\wr H_{q}$ and $G_{p}\times H_{q}$
are oligomorphic and transitive.

We first see that if $G_{p}$ is oligomorphic and transitive on $X_{p}$,
and if $H_{q}$ is transitive on $X_{q}$ with $q$ kept constant
with respect to $N$ (i.e. it is $\mathcal{O}(N^0)$) then $G_{p}\wr H_{q}$, $H_{q}\wr G_{p}$
and $G_{p}\times H_{q}$ are also oligomorphic and transitive. Note
that $G_{p}\wr H_{q}$ and $H_{q}\wr G_{p}$ are two different groups
in general. 

Furthermore, in the special case where both $G_{p}$ and $H_{q}$
are in oligomorphic and transitive families, then $G_{p}\wr H_{q}$,
$H_{q}\wr G_{p}$ and $G_{p}\times H_{q}$ are also in oligomorphic
and transitive families.

We can therefore construct the set of known oligomorphic and transitive
groups for any $N$. We proceed by recursive steps.
\begin{enumerate}
\item For some integer $n$, for $N=n$ the most basic subgroups of $S_{n}$
which are oligomorphic and transitive are $S_{n}$ itself and the
alternating group $A_{n}$ (as $A_{n}$ is $(n-2)$-transitive). We
have, a priori, 2 families.
\item For $N=n_{1}n_{2}...n_{d}$, we have the families $\bigotimes_{i=1}^{k}T_{n_{i}}$
and $T_{n_{1}}\wr T_{n_{2}}\wr...\wr T_{n_{k}}$ which are oligomorphic
and transitive for $T_{n_{i}}\in\{A_{n_{i}},S_{n_{i}}\}$ if $d\in\mathbb{N}$,
as the $T_{n_{i}}$'s are themselves transitive. We also include the
case $d=1$, where we just have $T_{N}$. We can extend this further
to all the possible combinations of wreath and direct products (taking
into account the fact that wreath products do not commute and that
both wreath and direct products are associative) of the $T_{i}$,
which we write $T_{n_{1}}\Box T_{n_{2}}\Box...\Box T_{n_{d}}$. We
write $\mathbb{A}_{N}=\{T_{n_{1}}\Box T_{n_{2}}\Box...\Box T_{n_{d}}|N=n_{1}...n_{d},d\in\mathbb{N},\Box\text{ a wreath or direct product}\}.$
\item Finally, for $N=n_{1}...n_{d}m_{d+1}...m_{2d+1}$, any group within
the ensemble $\mathbb{B}_{N}$ of known oligomorphic and transitive
families is of the form

\begin{equation}
    H_{m_{d}}^{(1)}\Box G_{n_{1}}^{(1)}\Box H_{m_{d+1}}^{(2)}\Box G_{n_{2}}^{(2)}...\Box H_{m_{2d}}^{(d)}\Box G_{n_{d}}^{(d)}\Box H_{m_{2d+1}}^{(d+1)},
\end{equation}
where $\Box$ is again either a direct or wreath product, and with the following restrictons:
\begin{enumerate}
\item $G_{n_{j}}^{(j)}\in\mathbb{A}_{n_{j}}$,
\item $H_{m_{j}}^{(j)}\leq S_{m_{j}}$ is either a transitive subgroup or
the trivial group if $m_{j}=1$. If it is a transitive subgroup, it
cannot be constructed by a product sequence involving any symmetric
or alternating groups (nor $A_{m_{j}}$ or $S_{m_{j}}$),
\item $d\in\mathbb{N}$, $m_{d+1}...m_{2d+1}\sim O(1)$. This necessarily
gives us that at least one of the $n_{i}$'s grow with $N$, avoiding
cases presented in the following remark. 
\end{enumerate}
\end{enumerate}
\begin{description}
\item [{Remark}] Note that we can also construct groups of the form $\bigotimes_{i=1}^{k}S_{n_{i}}$
and $\underset{k}{\underbrace{S_{n_{1}}\wr S_{n_{2}}\wr...\wr S_{n_{k}}}}$
for all $n_{i}\sim \mathcal{O}(1)$, and $k\sim \mathcal{O}(N)$. These are not
oligomorphic, as for a $K$-tuple, even for $K\ll N$, the number
of orbits is $\mathcal{O}(N)$.
For example, the group $\bigotimes_{i=1}^{\log_{2}N}S_{2}$ has size
$\left|\bigotimes_{i=1}^{\log_{2}N}S_{2}\right|=\left|S_{2}\right|^{\log_{2}N}=N$
and thus it cannot be oligomorphic. Likewise, $\underset{k}{\underbrace{S_{2}\wr S_{2}\wr...\wr S_{2}}}$
with $k=\log_{2}N$ has $\frac{1}{3}N^{\log3/\log2}$ orbits for $K=3$
tuples.
\end{description}
We have $\mathbb{A}_{N}\subset\mathbb{B}_{N}$, which will help us
to find $\left|\mathbb{B}_{N}\right|$. Note also that the construction
presented gives indeed non-conjugate subgroups, as for example for
all the constructions of the form $S_{n_{1}}\wr S_{n_{2}}...\wr S_{n_{d}}\wr H_{n}$
where $n\sim O(1)$ and the $n_{i}$'s all different, all different
ways to commute the product give back different conjugacy classes
since either the group order is different, or the group action is
different. 

We are now interested in calculating the lower bounds on the density
of $\mathbb{B}_{N}$. We will again proceed by steps: 
\begin{enumerate}
\item Let $c_{d}(N)$ be the number of possible non-conjugate subgroups
$T_{n_{1}}\Box T_{n_{2}}\Box...\Box T_{n_{d}}$ in $\mathbb{A}_{N}$
obtained by a factorization of $N=n_{1}...n_{d}$ into $d$ parts.
If we look at only the $S_{n_{1}}\wr S_{n_{2}}...\wr S_{n_{d}}$,
we see that we have as many such constructions as there are ordered
partitions of $N$ with $d$ parts. This gives a lower bound on $c_{d}(N)\geq f_{d}(N)$,
the number of ordered factorizations of $N$ of length $d$.
From \cite{Hwang2000,Lau2001,Sprittulla2016}, we
know that for $F_{d}(x)=\sum_{n\leq x}f_{d}(n)$, we have 
\begin{equation}
F_{d}(x)=x\frac{\left(\log x\right)^{d-1}}{\left(d-1\right)!}\left(1+O\left(\frac{d^{2}}{\log x}\right)\right).
\end{equation}
Since we are specifically interested in the case where $N\rightarrow\infty$
and $d$ finite, we can approximate this as $F_{d}(x)=x\left(\log x\right)^{d-1}/\left(d-1\right)!$,
and take the average value of 
\begin{equation}
f_{d}(N)\approx\frac{F_{d}(N)}{N}=\frac{\left(\log N\right)^{d-1}}{\left(d-1\right)!}.
\end{equation}
It is therefore clear that $f_{d}(N)$ is increasing with $N$

Note that there are also exactly $\left(\begin{array}{c}
d\\
i
\end{array}\right)$ ways to replace $i$ $S_{n_{j}}$'s by $A_{n_{j}}$'s in the product
$S_{n_{1}}\wr S_{n_{2}}...\wr S_{n_{k}}$. There is some subtlety
here, as we note that although $S_{n}\wr S_{n}$ commutes (and is
accordingly counted as a single factorization in $f_{d}(N)$), $S_{n}\wr A_{n}$
does not, hence both $S_{n}\wr A_{n}$and $A_{n}\wr S_{n}$ count
as two different terms. This gets fortunately caught by the combinatorial
terms.

Note also that direct products commute whilst wreath
products do not. Therefore the way to counts terms of the form $S_{n_{1}}\times...\times S_{n_{d}}$
is by counting the number of unorganized factorizations of length
$d$ of $N$, which we denote $u_{d}(N)$. Note that we can use the approximation $u_{d}(N)\sim \frac{1}{d!}f_{d}(N)$ For small $d$. The counting will however complicate once we allow a mixing of wreath and direct products. Indeed, $\left(A\wr B\right)\times C\neq A\wr\left(B\times C\right)$, hence we will get extra factors depending on the ways to arrange the product terms. Let us denote $C_{d}(N)$ the number of possible non-conjugate constructions $S_{n_{1}}\Box S_{n_{2}}\Box...\Box S_{n_{d}}$. We will therefore have
\begin{align}
    c_{d}(N) & =\sum_{i=0}^{d}\left(\begin{array}{c}
d\\
i
\end{array}\right)C_{d}(N) \\
& =2^{d}C_{d}(N)
\end{align}
From \cite{Sprittulla2016}, we have the relation
\begin{equation}
    f_{d}(N)=\sum_{i|N}^{N}f_{d-1}(N/i),
\end{equation}
with $f_{1}(N)=1$, which can help us out. Looking at the simple case where $d=3$, we have the following possible constructions, with associated counting
\begin{align*}
S_{n_{1}}\wr S_{n_{2}}\wr S_{n_{3}} & \rightarrow f_{3}(N)\\
S_{n_{1}}\wr\left(S_{n_{2}}\times S_{n_{3}}\right) & \rightarrow2!\sum_{\begin{array}{c}
n|N\\
n>1
\end{array}}u_{2}(N/n)\\
\left(S_{n_{1}}\wr S_{n_{2}}\right)\times S_{n_{3}} & \rightarrow\sum_{\begin{array}{c}
n|N\\
n>1
\end{array}}f_{2}(N/n)\\
S_{n_{1}}\times S_{n_{2}}\times S_{n_{3}} & \rightarrow u_{3}(N)
\end{align*}
Giving an overall counting of $C_{3}(N)=19/6\cdot f_{3}(N)-2f_{2}(N)$.

In general, let $T^{d}_{N}$ be a group of the form $S_{n_{1}}\Box S_{n_{2}}\Box...\Box S_{n_{d}}$. Then note that to construct a group of the form $T^{d+1}_{N}$, we have two systematic constructions with associated counting
\begin{align}
T^{m+1}_{n}\times T^{d-m}_{N/n} & \rightarrow \sum_{m=0}^{\left\lfloor\frac{d-1}{2}\right\rfloor}\sum_{\begin{array}{c}
n|N\\
n>1
\end{array}}C_{m+1}(n)C_{d-m}(N/n)\\
T^{m+1}_{n}\wr T^{d-m}_{N/n} & \rightarrow 2\sum_{m=0}^{\left\lfloor\frac{d-1}{2}\right\rfloor}\sum_{\begin{array}{c}
n|N\\
n>1
\end{array}}C_{m+1}(n)C_{d-m}(N/n).
\end{align}
Note that we have a slight overcounting for the cases where $T^{m+1}_{n}=T^{d-m}_{N/n}$ and the two therefore commute. This only happens if $d$ is odd and $n=\sqrt{N}$. We must therefore subtract an overall factor of $C_{(d-1)/2}(\sqrt(N))$ only if $d$ is odd. We can safely ommit it as we work in the large $N$ regime. Therefore, we have a reccurrence relation
\begin{align}
    C_{d+1}(N) & =3\sum_{\begin{array}{c}
n|N\\
n>1
\end{array}}\left[C_{d}(N/n)+\sum_{m=1}^{\left\lfloor\frac{d-1}{2}\right\rfloor}C_{m+1}(n)C_{d-m}(N/n)\right]\\
\Rightarrow C_{d+1}(N) & \sim 3a_{d}f_{d+1}(N)+\mathcal{O}\left((\text{log}N)^{(d-1)}\right),
\end{align}
where $C_{d}(N)=a_{d}f_{d}(N)+\mathcal{O}\left((\text{log}N)^{(d-2)}\right)$
 Therefore, for each length of factorization $d$, there are approximately
\begin{align}
c_{d}(N)\sim & \frac{6^{d}}{3}f_{d}(N)+\mathcal{O}\left((\text{log}N)^{(d-2)}\right)\\
= & \frac{6^{d}}{3}\frac{\left(\log N\right)^{d-1}}{\left(d-1\right)!}+\mathcal{O}\left((\text{log}N)^{(d-2)}\right)
\end{align}
possible non-conjugate subgroups. 
On the other hand, since $c_{d}(N)\geq f_{d}(N)$ as seen previously, we necessarily have that $c_{d}(N)$ scales as $(\log N)^{d-1}$
\item Now, the number of subgroups up to conugacy in $\mathbb{A}_{N}$,
which we denote $c_{\mathbb{A}_{n}}$ is given by adding up the $c_{d}(N)$'s
over all finite $d$'s. This therefore gives us 
\begin{align}
c_{\mathbb{A}_{n}} & \sim\sum_{d=1}^{l\sim O(1)}c_{d}(N) \sim\sum_{d=1}^{l\sim O(1)}\frac{\left(\log N\right)^{d-1}}{\left(d-1\right)!}6^{d}.
\end{align}
hence for arbitrary cutoff $l\ll N$, the number of elements in $\mathbb{A}_{N}$
is bounded from below by 
\begin{equation}
c_{\mathbb{A}_{N}}\geq \frac{6^{l}}{(l-1)!}(\log N)^{l-1}\label{eq:estimate of subgroup number for first construction}
\end{equation}
\item Now we can finally work out $c_{\mathbb{B}_{N}}$, the number
of subgroups up to conjugacy class in $\mathbb{B}_{N}$. We used the construction $H_{m_{1}}^{(1)}\Box G_{n_{1}}^{(1)}\Box H_{m_{2}}^{(2)}\Box G_{n_{2}}^{(2)}...\Box H_{m_{d}}^{(d)}\Box G_{n_{d}}^{(d)}\Box H_{m_{d+1}}^{(d+1)}$,
with $N=m_{1}n_{1}m_{2}n_{2}...m_{d}n_{d}m_{d+1}$ and $m_{1}m_{2}...m_{d+1}\sim O(1)$. We can already see that the contributions to $c_{\mathbb{B}_{N}}$ coming from varying the sizes and constructions of the $H_{m_{j}}^{(j)}$ will
be of finite order (given by the cutoff $l$). This will only change the prefactor in \ref{eq:estimate of subgroup number for first construction}. Hence we will have
\begin{align}
c_{\mathbb{B}_{N}}\sim & c_{\mathbb{A}_{N}} \\
\geq & (\log N)^{l-1}.
\end{align}
\end{enumerate}
Therefore, for any $l\ll N$, we have $c_{\text{oligo}\cap\text{trans}}(S_{N})\geq (\log N)^{l-1}$. We thus have the bounds
\begin{equation}
(\log N)^{l-1}\leq c_{\text{oligo}\cap\text{trans}}(S_{N})\leq e^{\frac{bN^{2}}{\sqrt{\log N}}},\label{eq:wreath and direct bound}
\end{equation}
for some $b$.

\begin{description}
\item [{Important Remark}] We will now discuss subgroups that are oligomorphic, but not transitive. We will see that the growth of such groups is parametrically faster, indicating that the transitivity condition is quite restrictive. First, note that we can form oligomorphic groups of
the type $S_{N-n}\times S_{n}$. We can define its action on two disjoint
sets $X_{N-n}$ and $X_{n}$, with the elements in $S_{N-n}\times S_{n}$
(which can just be seen as $g=(g_{1}g_{2})$ with $g_{1}$and $g_{2}$ acting
disjointly on $X_{N-n}$ and $X_{n}$). This group will have $2^{K}$
orbits when acting on $K$-tuples. In fact, for any $n\leq N/2$,
we have $S_{N-n}\times S_{n}$ having $2^{K}$ orbits. This family
of groups is therefore oligomorphic but not transitive. We can see
that all families of the type $S_{n_{1}}\times S_{n_{2}}\times...S_{n_{k}}$
with $N=n_{1}+n_{2}+...+n_{k}$, $k$ finite are of this type. We
denote this family $\mathbb{O}_{N}$.

When looking at groups in $\mathbb{O}_{N}$, we
see that for a given $N$, there are as many possible groups as there
are unordered partitions of $N$ into $N=n_{1}+n_{2}+...+n_{k}$,
$k$ finite, given by $p(k,N)$, with the asymptotics given by
\cite[Theorem 4.2.1]{Alfonsin2005}: 
\[
\sum_{k=1}^{l}p(k,N)=\frac{N^{l-1}}{l!(l-1)!}
\]
for $N\rightarrow\infty$ and $l$ a finite cutoff. We therefore see
that the number of families in $\mathbb{O}_{N}$ is bounded from below by $O(N^{l})$ for any finite $l$, which grows much faster than the
$O\left(\left(\log N\right)^{l}\right)$ for the families in $\mathbb{B}_{N}$.
Although this is not conclusive, it seems that the condition of transitivity
greatly reduces the number of available oligomorphic subgroup families
with which we can work to construct holographic CFTs.
\end{description}

\bibliographystyle{ytphys}
\bibliography{ref}

\end{document}